# Giant Magellan Telescope Site Testing Summary[*]

Joanna E. Thomas-Osip[†ab], Patrick McCarthy [ac], Gabriel Prieto [ab], Mark M. Phillips[bc], Matt Johns[a]
[a]Giant Magellan Telescope Organization, 813 Santa Barbara St, Pasadena, CA, USA 91101;
[b]Las Campanas Observatory, Colina El Pino S/N, Casilla 601, La Serena, Chile;
[c]Carnegie Observatories, 813 Santa Barabara St, Pasadena, CA, USA 91101;

## ABSTRACT

Cerro Las Campanas located at Las Campanas Observatory (LCO) in Chile has been selected as the site for the Giant Magellan Telescope. We report results obtained since the commencement, in 2005, of a systematic site testing survey of potential GMT sites at LCO. Meteorological (cloud cover, temperature, pressure, wind, and humidity) and DIMM seeing data have been obtained at three potential sites, and are compared with identical data taken at the site of the twin Magellan 6.5m telescopes. In addition, measurements of the turbulence profile of the free-atmosphere above LCO have been collected with a MASS/DIMM. Furthermore, we consider photometric quality, light pollution, and precipitable water vapor (PWV). LCO, and Co. Las Campanas in particular, have dark skies, little or no risk of future light pollution, excellent seeing, moderate winds, PWV adequate for mid-IR astronomy during a reasonable fraction of the nights, and a high fraction of clear nights overall. Finally, Co. Las Campanas meets or exceeds all the defined science requirements.

**Keywords:** site testing, precipitable water vapor, meteorological observations, seeing, turbulence profiles

## 1. INTRODUCTION

The GMT project is in the enviable position of having clear access to a developed site with a long history of excellent performance. The Las Campanas Observatory (LCO) in Chile has been known to be an outstanding site for nearly 40 years. The quality of the seeing at LCO is as good, or better, than at any other developed site in Chile, and weather patterns have been stable since the observatory initiated operations in 1971. There is negligible light pollution and little prospect of any in the future. After weighing both scientific and programmatic considerations, the GMT project, in concert with the GMT Board and Science Working Group, designated LCO as the baseline site and confined its site testing effort to characterizing sites within the LCO property. The aim was to identify the best peak within LCO in terms of seeing and wind speeds, and to determine the fraction of the time with PWV conditions suitable for mid-IR observations.

The Giant Magellan Telescope (GMT) site survey team has conducted an extensive study of potential GMT sites within the LCO property. A report detailing the properties of the sites, the techniques used to characterize their properties and the results has been prepared[1]. Here we summarize the essential aspects of the sites and the case that Co. Las Campanas is the appropriate site for the GMT. Further details can be found in Refs [1, 2, and 3].

## 2. OVERVIEW OF SITE TESTING PROGRAM

An extensive site testing program commenced in 2005 to identify the best available location at LCO for the GMT. Meteorological (cloud cover, temperature, wind, and humidity) and seeing data were obtained at three potential sites, and are compared with identical data taken at the Magellan telescopes site. The turbulence profile of the free atmosphere (above 500 m), precipitable water vapor, cloud cover and light pollution complement the site specific quantities.

The data from this program have been supplemented by the results from the Magellan site survey[4] (1988 – 1991), quantitative seeing information from Magellan and operator logs from the small telescopes. The Magellan survey measured meteorological data and a subset of the seeing parameters, but not PWV or turbulence profiles. Magellan records and operator logs provide long-term measures of the seeing and cloud cover. In this section, a description of the sites and instrumentation are presented.

---

[*] This paper includes data gathered with the 6.5 meter Magellan Telescopes located at Las Campanas Observatory, Chile.
[†] jet@lco.cl; phone +56-51-207-316; fax +56-51-207-308; www.lco.cl; www.gmto.org

## 2.1 Sites within the LCO property.

There are several developed telescope sites within LCO and there are a number of remaining sites with potential equal to those of the sites currently in use. Some of these will require more development than others. Four sites were monitored as part of the GMT site campaign and a subset of these are covered by the Magellan site survey data. The four sites are distributed along the ridge running NW to SE along the site and contain all sites considered suitable for a large telescope. These include, from NW to SE, Manquis Ridge and Cerros Manqui, Alcaino and Las Campanas, as is shown in Figure 1.

The Manquis Ridge site, between the du Pont and Swope telescopes, is the lowest altitude site with a large and reasonably level area along the ridge between these two telescopes. The Magellan site survey found it to have slightly poorer seeing than Co. Manqui but also lower wind speeds. Co. Manqui is home to the Magellan telescopes. Although there is not sufficient space to add another large telescope, it is the best characterized site at LCO and therefore it serves as a reference. Co. Alcaino, in between Cos. Las Campanas and Manqui and nearly the same altitude as Co. Manqui, was the site of the Nagoya 5-m radio telescope until 2004. It was not included in the Magellan site survey and the current studies provide the first detailed examination of its properties. A significant amount of rock would need to be removed to produce a large enough area for the GMT and associated facilities. Co. Las Campanas is the highest of our sites and was found by the Magellan site survey to have seeing similar to Co. Manqui but significantly higher wind speeds. Furthermore, although some rock will need to be removed at the summit, it has sufficient space for a large telescope and support facilities.

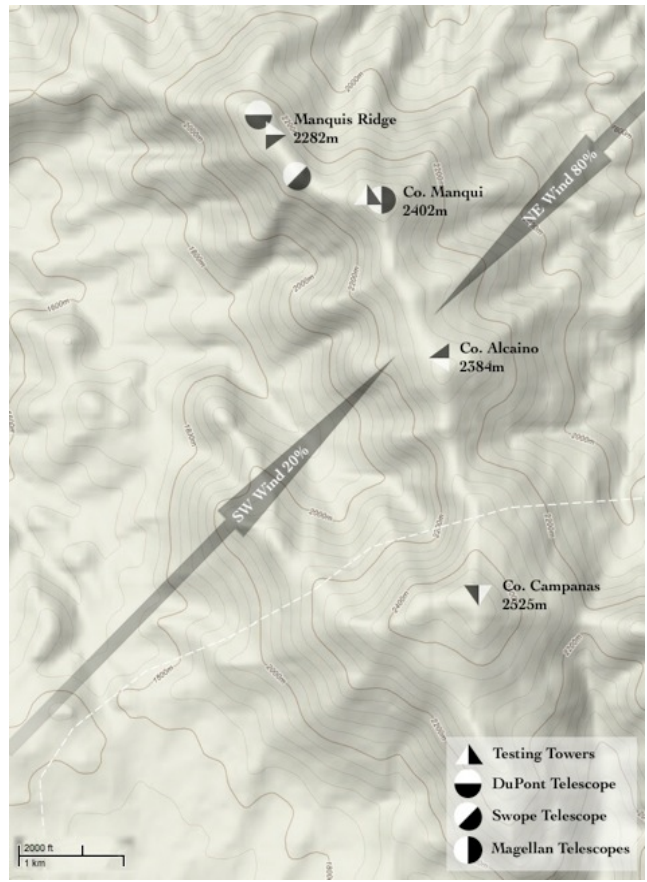

Figure 1. Sites tested at Las Campanas Observatory. The elevation of each site, in meters, is marked as are the prevailing and secondary wind directions. Co. Manqui is the Magellan site and serves as a reference point. DIMM instruments are mounted at each of the sites; the single MASS unit is deployed on Co. Manqui. The PWV monitor is located at Co. Alcaino. Co. Manqui (location of the twin Magellan Telescopes) is also shown. The towers that house the seeing instrumentation are also shown with their exact position on each site.

## 2.2 Instrumentation and Parameters Measured

Throughout the entire site testing effort nine different types of instruments (see **Table 1**) were used for a variety of purposes. Multiple copies of several instruments were used for monitoring site dependent quantities; bringing the instrument total to 17. All four of our sites had a seeing monitor (DIMM) and a weather station. Other quantities such as precipitable water vapor, cloudiness, and free atmosphere turbulence were assumed to be the same for all sites and the instruments were located for convenience.

Table 1. Site testing instruments with locations, dates in operation, and function.

| Instrument | Location(s) | Date | Parameters measured |
|---|---|---|---|
| Davis Instruments Corp Vantage Pro Weather Station | all | August 2005 -present | pressure, temperature, humidity, wind speed and direction |
| All Sky Camera[5] | Manquis Ridge | Apr 2005 - present | cloudiness |
| Tipping Radiometer[6] | Co Alcaino | Winter 2005 | PWV |
| IRMA[7] | Co. Alcaino | September 2007 – March 2010 | PWV |
| MIKE[6] | Clay telescope at Co. Manqui | July 2007 – March 2010 | PWV |
| DIMM[8] | all | November 2005 - present | seeing |
| MASS-DIMM[9] | Co. Manqui | September 2004 - present | free atmosphere Cn2 profiles, seeing, and AO parameters |
| Lunar Scintillometer LuSci[10] | Manquis Ridge | September 2007 and January 2008 | ground layer Cn2 profiles |
| SLODAR[11] | duPont telescope on Manquis Ridge | September 2007 and January 2008 | high resolution Cn2 profiles |

At each of the four sites two towers were erected, a 7-m tower for the seeing monitors and a 10-m tower for the weather stations. In some cases towers already existed from the Magellan site survey and only the domes were replaced. Each seeing tower consists of a concrete pedestal to which the telescope is bolted and an outer wind shield of culvert pipe that is not physically connected to the pedestal. Wind loads on the outer shield are carried by guide wires that are secured to the ground about 10-m away from the pedestal base. Observing floors for the operators to stand on are mounted to the wind shield and do not contact the pedestal. Computers and other electronic equipment that might produce heat and therefore dome turbulence were placed on a second floor below the observing floor. The towers all have aluminum 3-part clam-shell domes that open to reveal 2/3 of the full sky in a southerly direction to the monitors. Finally, all towers have a large extraction fan at the bottom that operates continuously during data collection to remove any heat produced in the tower.

# 3. METEOROLOGICAL STATISTICS

Fractional sky coverage by clouds is derived from visual observations made by the site test operators. Several times each night they record the fraction of the sky covered by clouds, on a scale from 0 (0%) to 9 (>90%). These visual estimates were checked and calibrated against a subset of quantitative images obtained with the CASCA all-sky camera. The results are summarized below.

Table 2. Cloud coverage statistics for LCO.

| Cloud Cover Index | Clear | ≤ 1/10 | ≤2/10 | ≤ 3/10 | ≤ 4/10 | ≤ 5/10 |
|---|---|---|---|---|---|---|
| Frequency | 0.64 | 0.69 | 0.75 | 0.80 | 0.84 | 0.87 |

One can deduce from **Table 2** that LCO it truly photometric approximately 64% of the time. The fraction of usable or "spectroscopic" time is more subjective but is in the range of 75%-80%. These fractions agree well with both the results of the Magellan site survey and the long-term logs from the small telescope operators. The cloud coverage has a strong seasonal component; the photometric fraction is highest in the summer months and falls to ~50% in the southern winter.

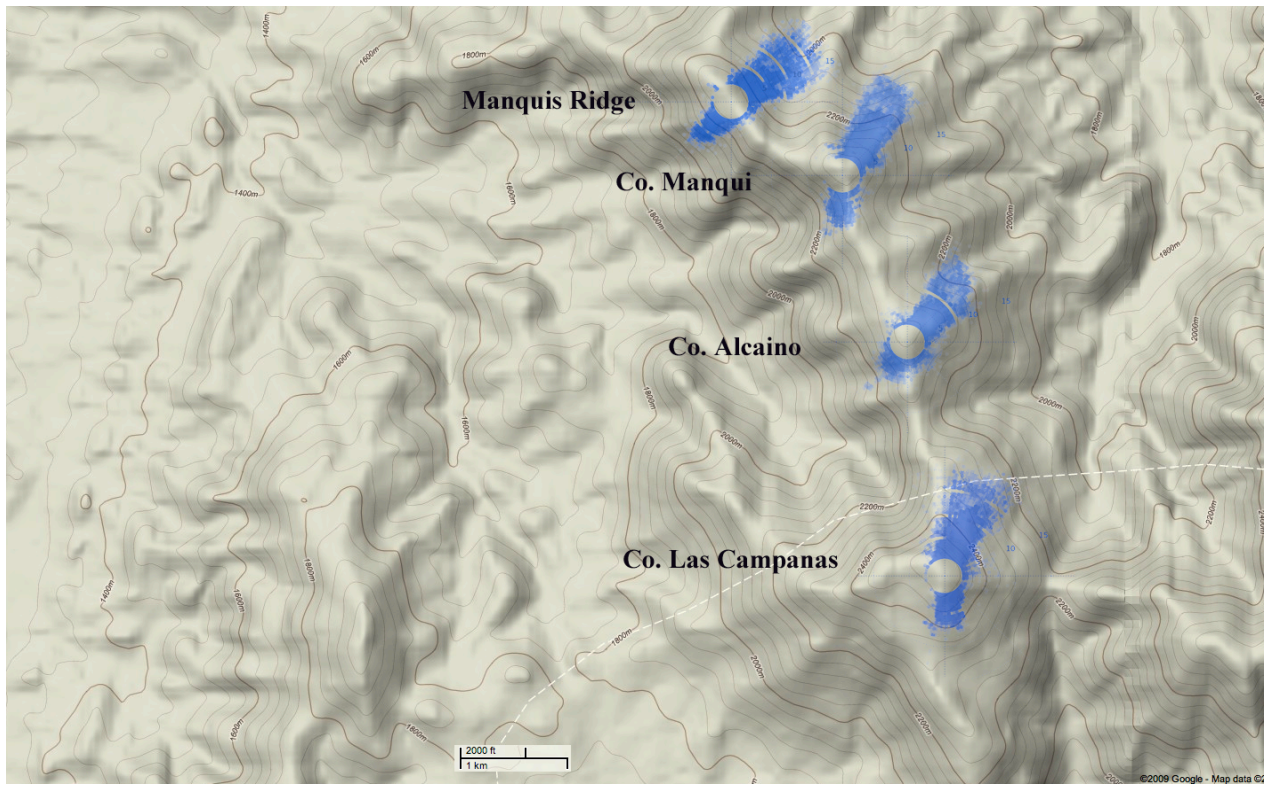

Figure 2. The wind roses for each of the four sites. The wind rose shows the amplitude and direction for each wind measurement. One can clearly see the bimodality of the wind direction and site-to-site variations due to local topography.

The wind at Las Campanas is highly regular. The majority of the time the wind is from the NE, approximately 20% of the time the wind comes from the opposite direction. Local topography modifies the wind directions slightly for the four tested sites and the wind speeds correlate with elevation. Co. Las Campanas, the highest site, has measurably stronger winds than the lower sites. In Figure 2, we summarize the qualitative aspects of the wind statistics by showing the wind roses (polar plot of direction versus speed) for the four sites superposed on a topographical map. From this one can see the bimodal distribution of wind directions and local topographic effects on the wind direction.

The median nighttime clear and partly clear weather wind speed (as displayed in Table 3) at the four sites ranges from 5.4m/s (Manquis Ridge) to 6.3m/s (Co. Las Campanas). The 95% percentiles under the same conditions range from 13.4 to 15.6 m/s. The current operating limit for the Magellan telescopes is 15.6 m/s, corresponding to 95% of clear/spectroscopic time at Co. Las Campanas. To meet the science requirement of less than 3% clear time lost to high winds, GMT should be designed to operate to a wind limit of 17.4 m/s (the 97th percentile). The wind statistics have a mild seasonal variation with the strongest winds found in late winter.

Table 3  Wind speed (m/s) percentiles (nighttime clear and partly clear)

|  | 10% | 25% | 50% | 75% | 90% | 95% | 97% | 99% |
|---|---|---|---|---|---|---|---|---|
| **Manquis Ridge** | 0.9 | 3.1 | 5.4 | 8.9 | 11.6 | 13.4 | 14.8 | 17.0 |
| **Manqui** | 1.3 | 3.1 | 5.8 | 9.4 | 13.0 | 14.8 | 15.6 | 17.9 |
| **Alcaino** | 1.3 | 2.7 | 4.9 | 8.0 | 11.2 | 13.0 | 13.9 | 15.6 |
| **Campanas** | 1.3 | 3.6 | 6.3 | 9.8 | 13.4 | 15.6 | 17.4 | 19.7 |

Temperatures at Las Campanas are quite benign and range in the extremes from -5ºC to ~ 25ºC, but the monthly median values range only from 7ºC to ~14ºC. LCO is a dry site in the sense that the relative humidity is low; the median values are around 35% and the winter months are considerably drier than the summer. The fraction of observing time lost to high humidity is less than 1% and the majority of closures due to humidity are less than 2 hours in duration.

One of the most important objectives of the GMT site testing campaign at LCO was to ascertain the PWV characteristics of the site since little information existed. Our campaign over a 2 year period covering two winter seasons shows (see Table 4) that the median column density is ~3.7 mm, while the best 10$^{th}$ percentile conditions correspond to ~1.2 mm of water. The winter months are dry and nights with PWV below 1mm are found 25% of the time. Finally, our data show that PWV is typically stable over the period of a night. For further details see Ref [2].

Table 4.  Clear nighttime calibrated IRMA PWV (mm) percentiles.

| Season | 10% | 25% | 50% | 75% | 90% | % < 1.5 mm | Samples |
|---|---|---|---|---|---|---|---|
| **All** | 1.2 | 2.1 | 3.7 | 6.1 | 8.2 | 15 | 186300 |
| **Winter** | 0.5 | 0.9 | 1.4 | 2.0 | 2.7 | 55 | 13312 |
| **Spring** | 1.0 | 1.4 | 2.1 | 3.2 | 4.2 | 28 | 58594 |
| **Summer** | 2.0 | 3.0 | 5.1 | 7.1 | 10.0 | 4 | 48633 |
| **Fall** | 2.9 | 3.7 | 4.8 | 6.6 | 8.2 | 3 | 65761 |

## 4. IMAGE QUALITY STATISTICS

The seeing at each the four sites has been measured using a differential image motion monitor (DIMM). Free atmosphere seeing is monitored with a multi-aperture scintillation sensor (MASS) at one of the sites. The overall seeing characteristics of the four sites are excellent and the GMT site monitoring results are in close agreement with the results from the Magellan site survey and measurements from the Magellan guide cameras. The statistics of the overall seeing for the four sites are summarized in Figure 3 and Table 5.

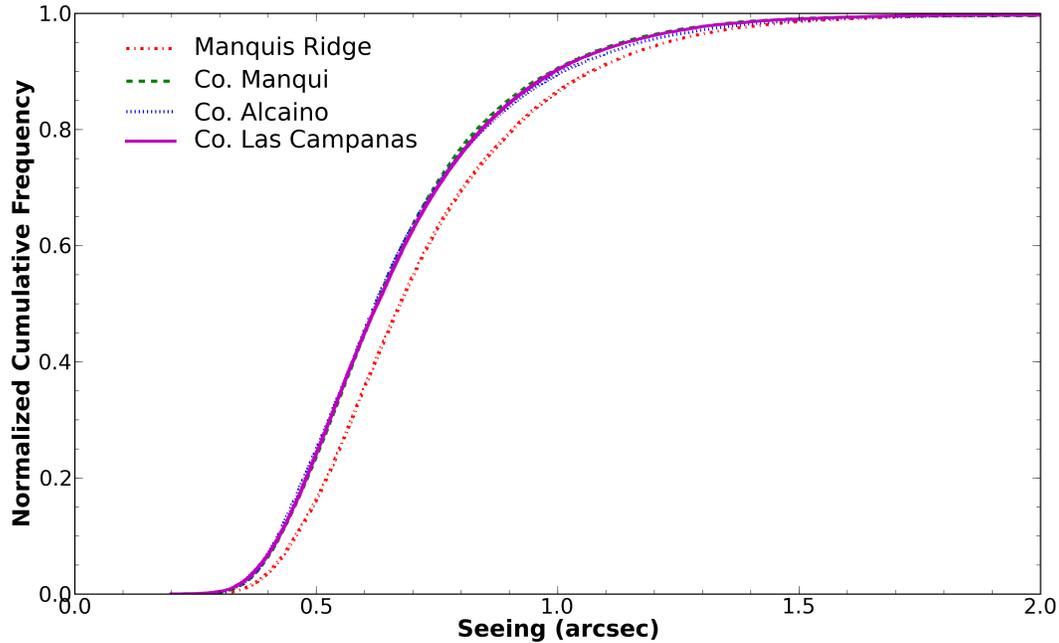

Figure 3. Histogram of DIMM seeing for the four sites. One can clearly see that the seeing at the Manquis Ridge site is inferior.

The median seeing at the four sites ranges from 0.62 to 0.67" FWHM. The DIMM seeing at Co. Las Campanas, the Magellan site (Co. Manqui) and Co. Alcaino are indistinguishable. The seeing at Manquis Ridge is clearly inferior to that at the other sites, as was determined during the Magellan site survey. Furthermore, good seeing stability (analyzed by examining the durations of periods with seeing below 0.6") is also inferior at Manquis Ridge.

Table 5. Seeing statistics for the four sites.

| Seeing Percentiles | 10% | 25% | 50% | 75% | 90% |
|---|---|---|---|---|---|
| Manquis Ridge | 0.46 | 0.55 | 0.67 | 0.85 | 1.07 |
| Co. Manqui | 0.42 | 0.51 | 0.62 | 0.79 | 0.99 |
| Co. Alcaino | 0.42 | 0.5 | 0.62 | 0.79 | 1.01 |
| Co. Las Campanas | 0.42 | 0.5 | 0.63 | 0.79 | 0.99 |

The strength of the ground layer and the relative contributions of the free atmosphere and ground layer can be deduced from a number of techniques. We have used several approaches to gauge the properties of the ground layer. These include MASS/DIMM comparisons, SLODAR, and lunar scintillation[12,13]. Each approach has its particular strengths and, in a broad sense, the various techniques give results that are in agreement. The largest data set that we have relating to ground layer is the combined MASS/DIMM measurements. These give a measure of the seeing in the lowest ~500 m of the atmosphere and the seeing for all turbulence layers above this elevation. In Table 4 we summarize the properties of the ground layer above the four sites based on our analysis of the MASS/DIMM data. A significant fraction of the seeing (~65%) can be attributed ground layer turbulence above the site in median conditions.

Table 6. Seeing in the ground layer and free atmosphere, in arcseconds FWHM, at each of the four sites. The thickness of the ground layer sensed based on the altitude of the MASS instruments and each individual DIMM is listed in the rightmost column.

| Seeing Percentile | 10% | 25% | 50% | 75% | 90% | ground layer thickness (m) |
|---|---|---|---|---|---|---|
| Manquis Ridge | 0.20 | 0.31 | 0.44 | 0.59 | 0.79 | 642 |
| Co. Manqui | 0.18 | 0.27 | 0.37 | 0.50 | 0.67 | 500 |
| Co. Alcaino | 0.17 | 0.27 | 0.38 | 0.52 | 0.72 | 540 |
| Co. Las Campanas | 0.17 | 0.27 | 0.38 | 0.52 | 0.71 | 399 |
| MASS free atmosphere | 0.24 | 0.32 | 0.45 | 0.63 | 0.85 | - |

Average free atmosphere turbulence profiles from the MASS indicate that in poor seeing conditions most of the turbulence is found below 4 km and the opposite is true in good seeing conditions. Furthermore, when the average profiles are sorted by ground layer seeing there is very little distinction between good and bad conditions supporting the idea of an independent ground layer and free atmosphere. A two dimensional histogram of the free atmosphere and ground layer seeing, displayed in Figure 4, shows the most common conditions for both the free atmosphere and ground layer seeing to be ~0.3". Furthermore, we find a strong cut-off in the free atmosphere seeing near 0.1" while the ground layer seeing shows almost no lower cut–off. The lowest ground layer seeing values are found when the free atmosphere is above 0.4". This is likely an instrumental effect due to the fact that the ground layer seeing is calculated as a difference of two often similar values.

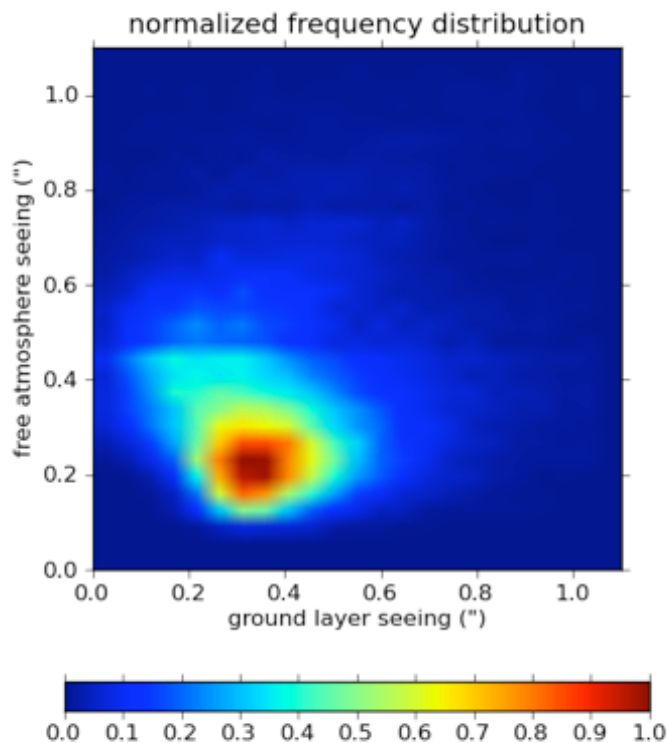

Figure 4 Image showing the two dimensional normalized histogram of the ground layer and free atmosphere seeing at Cerro Manqui.

There appears to be a mild seasonal variation in the seeing (as shown in Figure 5) with the summer and fall months having better seeing. Good seeing conditions (e.g. < 0.5" FWHM), however, occur at all times of the year. The seasonal variation in the total seeing is due to a seasonal variation in the free atmosphere seeing as no variation is seen in the ground layer seeing. There are also weak correlations between seeing and wind speed, with high winds having poorer seeing, but the effect is not large and is only apparent at high wind speeds. None of the sites show a correlation between seeing and wind direction. Furthermore, the full atmosphere turbulence profiles are also weakly correlated with ground wind speed in the same sense as the seeing. Many more details of the seeing and turbulence profile statistics can be found in Ref [3] in these proceedings.

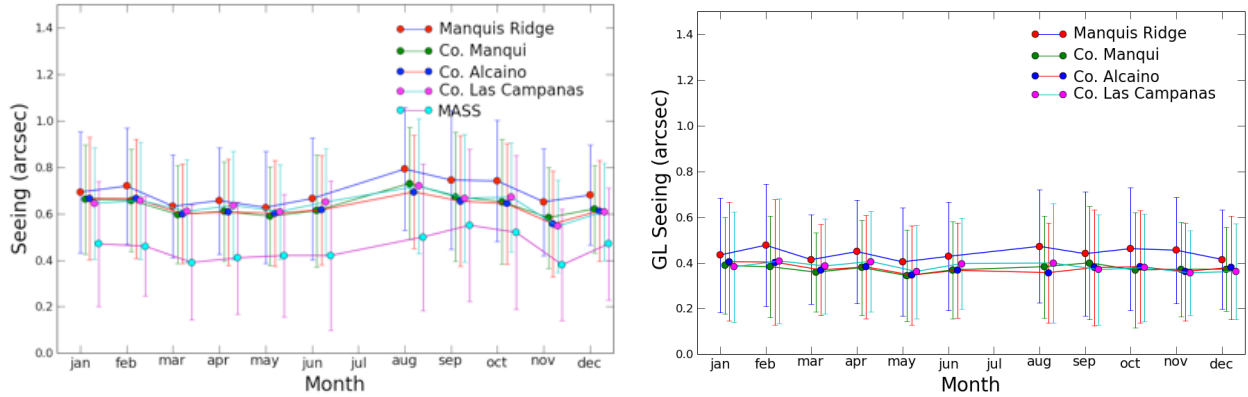

Figure 5 Right: Median seeing at all four sites and in the free atmosphere (MASS) as a function of month. Left: Median seeing in the ground layer (GL) at the four sites as a function of month. This plot includes 3 years of data. The errors correspond to the standard deviation of the data within the month. For clarity, the data for each site are offset by 0.1 months from each other. The data for July have been removed due to the low number of data points in that bin.

## 5. THE CASE FOR CERRO LAS CAMPANAS

The long history of excellent conditions at LCO, concerns regarding biases in short-term site surveys, and limited resources led the GMT project to focus its site evaluation activities to certain peaks within the LCO property. LCO has dark skies, little or no risk of future light pollution, excellent seeing, moderate winds, PWV characteristics that will meet the science goals, and a high fraction of clear nights.

Table 7 compares the GMT site requirements and goals as specified in the GMT Conceptual Design Review[14]. As may be seen, in nearly all of the parameters, each of the four sites tested at LCO meet the requirements, and in several the goals are met or exceeded. Three of the four sites that have been characterized, Cos. Manqui, Alcaino and Las Campanas have excellent and very similar properties. Co. Las Campanas and the Magellan site (Co. Manqui) are nearly identical in their seeing statistics and, apparently, their average ground-layer characteristics. The only significant difference between the two sites is the stronger wind at Co. Las Campanas. If the same operating limits are employed, this would result in a small increase in the amount of time lost to wind. The greater wind speeds have a positive aspect as they mitigate periods of poor seeing due to low winds and ineffective flushing of warm air in the enclosure and in the ground layer.

Table 7  Comparison of GMT Site Requirements with Site Testing Results

| Property | Requirement (Goal) | Manquis Ridge | Co. Manqui | Co. Alcaino | Co. Las Campanas |
|---|---|---|---|---|---|
| **Median FWHM Seeing (")** | <0.65 (≤0.5) | 0.67 | 0.62 | 0.62 | 0.63 |
| **Clear (%)** | >60 (>70) | 64±4 | | | |
| **Clear + Partly Clear (%)** | >70 (>80) | 80±4 | | | |
| **Wind Speed > 15.6 m/s (%)*** | <3 | 2 | 3 | 1 | 5 |
| **Percentile with PWV < 1.5 mm** | 10th (15th) | 15th | | | |

*The current Magellan wind limit is 15.6 m/s while GMT will need a limit of 17.4 m/s to meet the science requirement.

The data summarized above show that both Cos. Las Campanas and Alcaino sites are similar in their properties and quite similar to the Magellan site. Practical considerations concerning the site area and profile of the mountaintop must be taken into account in making a site selection. Co. Alcaino is rather small and sharply peaked.  If the site were cleared to accommodate the GMT enclosure and support building little of the peak would remain (see Figure 6). It is not clear that the site would retain its properties. There is some concern that it would resemble the Manquis Ridge site in its topography after clearing and could well mimic the poor seeing at that site.

Co. Las Campanas, however, has far greater room at the summit. While a significant amount of material would need to be removed to accommodate the GMT, the site would retain its essential peak structure and therefore has the best layout for the accommodation of a large telescope. For this reason we believe that practical considerations strongly favor Co. Las Campanas over Co. Alcaino.

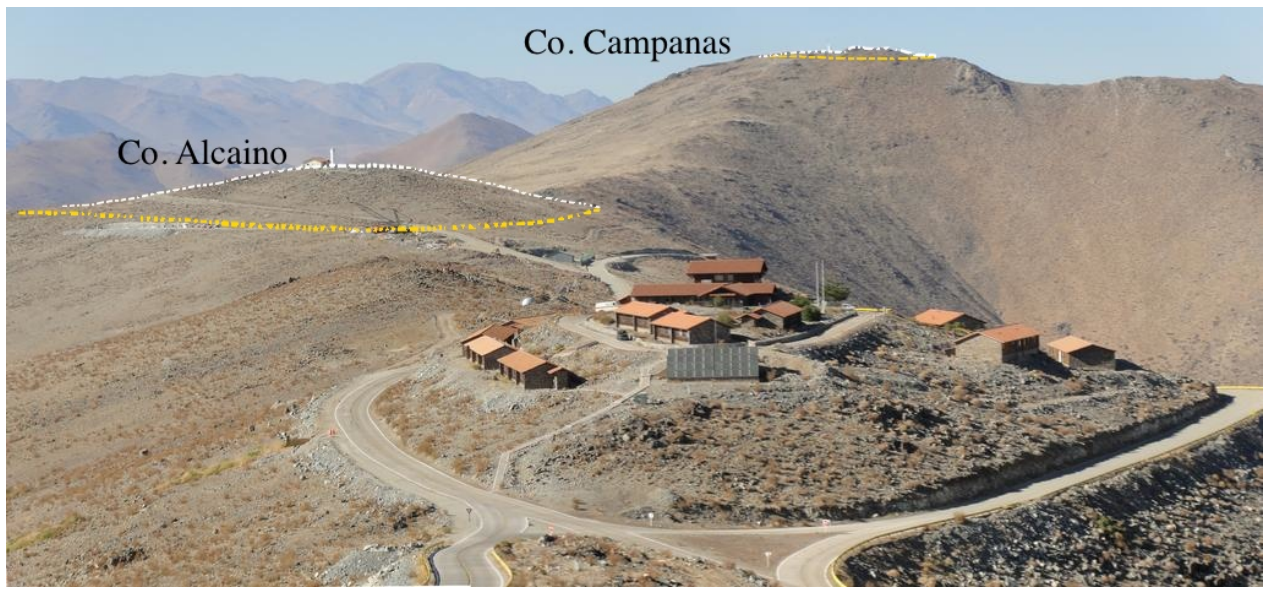

Figure 6. Photo of Cos. Alcaino (foreground) and Las Campanas (background). The amount of earth removal necessary for GMT is indicated in yellow

The site survey team and the GMT Project firmly believe that Co. Las Campanas is the optimum site for the GMT and it has been formally designated as the GMT site.

## AKNOWLEDGEMENTS

We would like to acknowledge the contributions of Alex Athey, Christoph Birk, Pablo Castro, Emilio Cerda, Felipe Daruich, David Floyd, Javier Fuentes, Gaston Gutierrez, Andrew McWilliam, Cesar Muena, Frank Perez, Myriam Perez, Miguel Roth, Steve Shectman, Josefina Urrutia, Sergio Vera and the rest of the LCO staff. Finally, we are indebted to the many Magellan MIKE observers.

This material is based in part upon work supported AURA through the National Science Foundation under Scientific Program Order No. 10 as issued for support of the Giant Segmented Mirror Telescope for the United States Astronomical Community, in accordance with Proposal No. AST- 0443999 submitted by AURA. Any opinions, findings, and conclusions or recommendations expressed in this material are those of the author(s) and do not necessarily reflect the views of the National Science Foundation.